\documentstyle[12pt]{article}
\newcommand{\NP}{{Nucl.\ Phys.\ }}
\newcommand{\PL}{{Phys.\ Lett.\ }}

\newcommand{\PRL}{{Phys.\ Rev.\ Lett.\ }}

--
\begin{document}
\pagestyle{empty}
\begin{flushright}
UCT-TP 212/94  \\
hep-th/9405103
\end{flushright}
\vspace*{31mm}
\begin{center}
{\large \bf
SUPERCOOLING OF THE QUARK-GLUON PLASMA\footnote
{Talk at the XXIXth RENCONTRE DE MORIOND,
 "QCD and High Energy Hadronic Interactions",
 Meribel, France, March 19-26, 1994}
}\\[24pt]
{\bf Neven Bili\'c} \\
Department of Physics,
University of Cape Town, Rondebosch 7700, South Africa  \\
and \\
Rudjer Bo\v{s}kovi\'{c} Institute,
P.O. Box 1016, Zagreb, Croatia                          \\
\vskip 6cm
{\bf ABSTRACT }      \\
\end{center}
 Transition from the quark-gluon (QG) plasma
 to a hadronic gas is studied in the framework of
 the relativistic combustion theory.
Energy-momentum conservation and baryon number conservation
 constrain the possible temperature jump across the front
 separating the two phases.
Assuming the
  temperature in the hadronic phase to be fixed
  from experiment
  one can
determine the corresponding temperature in the QG phase.
The calculations reveal
that the QG phase must be in a strongly supercooled state.
The stability of this solution with respect to minor
modifications is investigated. In particular
the effect of an admixture of hadronic matter in the QG phase
 is considered in detail.
 By increasing the fraction of hadronic matter the supercooling
becomes
 weaker and eventually the transition proceeds via a smoother
 deflagration wave.
\newpage
\section{Introduction}
The principal goal of high energy heavy ion collisions is to produce
the
QG plasma.
The plasma formation and its subsequent disintegration is believed
 to proceed in three steps \cite{muller}
 pre-equilibrium equilibrium
 and hadronisation.
 Attempts to describe the hadronisation as an equilibrium
  transition from the QG plasma to a hadronic gas
 at certain temperature and baryon chemical potential
 face the following problems:
First, in order to accommodate the existing
experimental data
\cite{NA35,WA85,NA36,helstrup},
the transition temperature must be $T\simeq$ 200 MeV
and the baryon chemical potential
200$<\mu_B<$300 MeV
\cite{satz1}-\cite{rafelski1}.
This is in disagreement with
 lattice calculations  predicting the critical temperature
  of the order or lower than 150 MeV even for
$\mu_B$=0.
Second, the entropy per baryon in the QG phase is greater than
in the hadronic phase so that the transition would violate the
second law of thermodynamics.
As a first attempt to cure some of the problems we
propose a model based on
 a relativistic
combustion of the QG plasma into a hadronic gas.
 Assuming that the
 thermodynamic properties of the hadronic gas are fixed by
 experiment one can determine the properties of QG plasma
 by making use of various conservation laws.
\section{Equation of State }
The QG plasma is basically an ideal gas of quarks and gluons
in which the influence of the QCD vacuum is implemented
either by subtracting the bag constant $B$ from the pressure
 \cite{oertzen}
or introducing a low momentum
 cut-off, motivated by numerical results
from lattice gauge theories \cite{karsch}.
In the cut-off model
the density of gluons, for example, is given by
\begin{equation}\label{eq1}
n=16\int_{k_c}^{\infty} {d^3p\over (2\pi )^3}{1\over e^{E/T}\pm 1}
\end{equation}

We have chosen a value of $5T_c$ for the cut-off
parameter in order to reproduce the critical temperature
$T_c\simeq 150$MeV
obtained in lattice gauge theories for
$\mu_B$=0 \cite{gottlieb}.
 Certain quantities are very sensitive to the precise value of
this cut-off and to show this influence we also present for comparison
results obtained with a value of
$3.5T_c$.

In order to study the effects of possible hadronic bubble
formation we shall consider
that the QG phase was not pure but that some  admixture of
hadronic matter exists
consisting for simplicity of only pions \cite{redlich}.
Thus for example, the energy density is given by
\begin{equation}\label{eq5}
\epsilon_q=(1-f)\epsilon_q^0+f\epsilon_{\pi}
\end{equation}
where $\epsilon_q^0$ refers to the pure QG matter
and $f$ is the fraction of hadronic matter present in the QG
plasma phase.
Similar expressions hold for
the pressure $P_q$, the
entropy density $s_q$
and the baryon number density $n_q$.

The hadronic side in our approach is a composition of non-interacting
quantum gases.
We include  all
well established hadronic resonances
 \cite{particles}.
We incorporate the hard core radius of hadrons
 \cite{helmut}
by choosing
a typical proper volume of a hadron with a radius of
 0.8 $fm$.
Since the chemical potential $\mu_S$
       is fixed by the strangeness neutrality
      condition \cite{satz1}
 the thermodynamical quantities are functions of
 $T$ and $\mu_B$ only.
\section{Relativistic Combustion}
Let us assume that a sharp boundary
 (front) exists separating the two phases and that
both sides of the front are  in thermal equilibrium
at different temperatures and
chemical potentials.
Energy and momentum conservation across the front
leads to the
 equations
 \cite{landau} :
\newpage
\begin{equation}\label{eq3}
(\epsilon_h + P_h)\gamma_h^2 v_h + P_h =(\epsilon_q + P_q)\gamma_q^2
 v_q+P_q ,
\end{equation}
\begin{equation}\label{eq4}
(\epsilon_h+P_h)v_h^2\gamma_h^2=(\epsilon_q+P_q)v_q^2\gamma_q^2 ,
\end{equation}
The index $q$ refers to the QG plasma side
which has an admixture of hadronic matter, while the index $h$
refers to the hadronic phase. $v$ is the velocity
of the gas in each phase with
respect to the front, and $\gamma =
1/\sqrt{1-v^2}$.
In addition to the above two equations, baryon number conservation
 leads to
\begin{equation}\label{eq6}
n_hv_h\gamma_h=n_qv_q\gamma_q.
\end{equation}

In order to determine the four unknown quantities
$\epsilon_q, P_q, v_q$ and $v_h$ as functions of
$\epsilon_h, P_h$ and $n_h$ we need one more equation.
This is given by the requirement of non-decreasing entropy
which implies in our notation
\begin{equation}\label{eq7}
s_hv_h\gamma_h\geq s_qv_q\gamma_q \quad ,
\end{equation}
As has been shown previously \cite{oertzen}, eqs.
(\ref{eq3}-
\ref{eq7}) yield a
supercooled QG plasma.
The equality sign in (\ref{eq7}), corresponding to an adiabatic
process,
determines the minimal amount of supercooling needed to fit
the hadronic data without violating the second law of
thermodynamics.
 Combining (\ref{eq6}) and the equality in (\ref{eq7}) we find
\begin{equation}\label{eq9}
{s_h\over n_h}={s_q\over n_q} ,
\end{equation}
which may be used
as the fourth constraint, keeping in mind that the QG
temperature $T_q$ thus obtained is maximal.

Equations
(\ref{eq3},
\ref{eq4},
\ref{eq6})
 can be solved for the velocities, leading to
\begin{equation}\label{eq10}
v_q^2={(P_h-P_q)(\epsilon_h+P_q)\over
(\epsilon_h-\epsilon_q)(\epsilon_q+P_h)};
\; \; \;
v_h^2={(P_h-P_q)(\epsilon_q+P_h)\over
(\epsilon_h-\epsilon_q)(\epsilon_h+P_q)} .
\end{equation}

\begin{minipage}[t]{98mm}
\vspace*{91mm}
\begin{center}
\parbox{77mm}
{\footnotesize Fig.~1.
Energy-density of the hadronic gas (lower curve)
and of the QG plasma (upper curve) as a function
of temperature. Super-cooling and super-heating are
indicated by the short-dashed line.}
\end{center}
\vspace*{1mm}
\end{minipage}
\begin{minipage}[t]{7cm}
With help of eq. (\ref{eq6}) the velocities can be eliminated
yielding
\begin{equation}\label{eq12}
\left({n_q\over n_h} \right)^2=
{(\epsilon_q+P_q)(\epsilon_q+P_h)\over
(\epsilon_h+P_h)(\epsilon_h+P_q)}.
\end{equation}
This equation is equivalent to Rankine-Hugoniot-Taub adiabat
equation \cite{thorne}.
Combining now equations (\ref{eq9}) and (\ref{eq12}) we can calculate
numerically
the temperature $T_q$ and the chemical potential $\mu_B^q$.
This in turn yields all thermodynamic quantities on the
QG side.
In Fig. 1 we show our results for the case when there is no admixture
of
hadronic matter in the QG plasma . There is a substantial
amount of supercooling in the QG phase. The transition then takes
place
to a hadronic gas which has a temperature $T= 200$ MeV as suggested by
most experiments at CERN using a relativistic ion beam.
\end{minipage}

\newpage
Most results from
lattice gauge theories indicate that this temperature is higher than
the
critical temperature of the phase transition. We thus  conclude
that the hadronic gas is in a super-heated phase.

This scenario immediately raises the following question : are there
mechanisms preventing this strong super-cooling? In particular it has
been argued \cite{kapusta} that super-cooling
enhances the formation of hadronic bubbles in the QG plasma.
 We have therefore considered admixing hadronic matter to the
QG plasma phase.

\begin{minipage}[t]{98mm}
\vspace*{90mm}
\begin{center}
\parbox{77mm}
{\footnotesize Fig.~2.
The quark temperature $T_q$ versus the
fraction of hadronic admixture $f$ in the QG phase for
  cut-off values of $5T_c$ (solid line) and
 $3.5T_c$ (dashed line).}
\end{center}
\vspace*{1mm}
\end{minipage}
\begin{minipage}[t]{7cm}
The results are shown in Fig.~2.
Keeping the parameters on the
hadronic side fixed at $T=200$ MeV and $\mu_B=300$ MeV we vary the
fraction of hadronic admixture $f$ in th QG plasma as
introduced in equation (\ref{eq4}), and
 we calculate each time
the resulting value of the temperature on the QG side of the
front.
 If the cut-off value is large
$5T_c$ then there is an abrupt turnover when the admixture exceeds
20\%.
The temperature increase across the front is reduced after this.
However
when the cut-off is taken at $3.5T_c$
or
if a bag-model equation of state is used
 then the picture is completely
different : the strong supercooling persists anomalously long until
the
admixture has reached more than 60\%, in which case it is no longer
possible to speak of a ``small'' admixture.
\end{minipage}

It is customary to work in the frame where the sharp front separating
the
two phases is at rest.
In this frame the QG plasma moves towards the front with a
velocity $v_q$ and the hadrons move away from the front with a
velocity
$v_h$.
If $v_q<v_h$ one has a deflagration, if $v_q>v_h$ one has a detonation
 \cite{Vanhove}-\cite{esko1}.
As is well known, small perturbations in a medium  propagate with the
velocity of sound in that medium. It is therefore of special interest
to
compare $v_q$ and $v_h$ with the velocity of sound in each phase.
The latter is defined as \cite{landau}
\begin{equation}\label{eq14}
c_s^2=\left.{\partial P\over\partial\epsilon}\right|_{s/n}=
{\partial P/\partial T +\partial P/\partial\mu_B\cdot
d\mu_B/dT
      \over
 \partial\epsilon/\partial T +\partial\epsilon/\partial\mu_B\cdot
d\mu_B/dT} ,
\end{equation}
where the  derivative
$d\mu_B/dT$ is taken at constant $s_h/n_h$ .
A detonation is classified as strong if
$v_q>c_{s,q},\; v_h<c_{s,h}$,
 whereas a deflagration is strong if
$v_q<c_{s,q},\; v_h>c_{s,h}$ \cite{courant}

One could, in principle, experimentally distinguish between
detonation and deflagration by measuring the transverse flow of
produced hadrons and compare with hadronic velocities $V_h$ calculated
in the laboratory frame, in which QG plasma is at rest.
One finds for a detonation and a deflagration respectively
\begin{equation}\label{eq17}
V_{h,det}=\frac{v_q +v_h}{1+v_q v_h},
\; \; \;
V_{h,def}=\frac{v_q -v_h}{1-v_q v_h}.
\end{equation}

\newpage
The velocities
$v_q^2$ and $v_h^2$
given by (\ref{eq10})
and sound velocities calculated using (\ref{eq14})
  are depicted in Fig.~3
as functions of the fraction $f$.
The change
in the transition mechanism leads to very sharp variations in $v_q$
and $v_h$.
 The velocities  are unphysical  between the
points A and A'. In the region of strong detonation
(up to the point O) the laboratory hadronic velocity
$V_{h,det}$ is in the range 0.6 - 0.7.
In the strong deflagration region (point O' onwards)
$V_{h,def}$ ranges between 0.2 - 0.5.
Thus, a deflagration is characterized by a substantially
lower transverse flow.A direct comparison with
experiment, however, would be possible only after solving the
equations of motion in a more realistic
(radial or cylindrical) geometry.
\vspace*{100mm}
\begin{center}
\parbox{130mm}
{\footnotesize Fig.~3.
The velocity squared of the  QG matter $(a)$
and the hadronic matter $(b)$  in the rest-frame
of the front.
The dashed line is the velocity of sound squared in the
corresponding medium.
The value of the
cut-off parameter is $5T_c$.}
\end{center}

\section{Conclusions}
In this paper we have studied the consequences arising from the
existence of a sharp front separating the QG plasma phase
from
the hadronic phase. The equations of continuity for energy, momentum
and
baryon number as well as the constraint of nondecreasing
entropy, require  a substantial amount
of
supercooling to occur in the QG plasma before the transition
can take place. This transition then leads to a super-heated
hadronic gas. We have also investigated what happens if the QG
plasma is not pure but contains an admixture of hadronic matter.
 We found that from the  moment when the admixture
of
hadronic matter is more than about 20\% of the total
 then the
transition
proceeds via a
deflagration wave
 In this case the transition is  smoother and the temperature jump
across the front is not as pronounced as in the absence of
admixture of hadronic matter.

 The crossover region between the two solutions shows a remarkable
 S-shape structure. The multiple valued $T_q$ and $\mu_B^q$
 in that region indicate that in the given range of $f$
 a chaotic turbulent type of transition could take place.

%
\newpage
{\large \bf Acknowledgement}

This paper is based on a work done in collaboration with
J. Cleymans, D.W. von Oertzen, \\ K. Redlich and E. Suhonen.
\end{document}